\documentclass[3p,times,twocolumn]{elsarticle}
\usepackage{ecrc}
\volume{00}
\firstpage{1}
\journalname{Nuclear Physics B Proceedings Supplement}
\runauth{}
\jid{nuphbp}
\jnltitlelogo{Nuclear Physics B Proceedings Supplement}

\usepackage{amssymb}

\usepackage[figuresright]{rotating}

\def\ccqe {CC~QE }
\def\ccpip {CC~1$\pi^+$ }

\def\numuccqe {$\nu_{\mu}n\rightarrow\mu^-p$}

\def\numuccpip {$\nu_{\mu}N\rightarrow\mu^-N'\pi^+$}
\def\maqe {$M_A^{QE}$}
\def\enu {$E_{\nu}$}
\def\q2 {$Q^2$}
\def\gev2 {(GeV/c)$^2$}
\def\numu {$\nu_{\mu}$}

\def\mup {$\mu\!-\!p$ }
\def\mupi {$\mu\!-\!\pi$ }

\begin{document}

\begin{frontmatter}


\title{Recent Measurements of Neutrino-Nucleus Quasi-Elastic Scattering}
\author{M.~O.~Wascko}
\ead{m.wascko@imperial.ac.uk}
\address{Imperial College London, Physics Department, London SW7 2BW, United Kingdom}

\begin{abstract}

We present recent measurements of neutrino charged current
quasi-elastic (\ccqe) scattering, \numuccqe .  Measurements of \ccqe
on carbon near 1~GeV by MiniBooNE and SciBooNE, as well as
measurements on iron at 3~GeV by MINOS, disagree with current
interaction models, while measurements at higher energies on carbon by
NOMAD show excellent agreement with those same models.

\end{abstract}


\end{frontmatter}


\section{Introduction}

Neutrino physics is entering a new era of precision measurements of
oscillation parameters.  The measured value of the atmospheric
neutrino mass splitting is such that current and future accelerator
neutrino beams are best tuned to oscillation physics with neutrino
energies in the few-GeV region.  However, the precision of neutrino
interaction cross-sections is not commensurate with the goals of the
next generation of neutrino oscillation
experiments~\cite{Itow:2002rk,Harris:2004iq}. Moreover, recent
measurements have exposed serious shortcomings in the current
theoretical models describing neutrino-nucleus interactions.

One of the largest interaction processes in the few-GeV region is
quasi-elastic scattering (\ccqe ), \numuccqe.  The \ccqe process is
important because it is the signal reaction for oscillation
experiments with neutrino energies below $\sim$2 GeV and because the
simple final state allows accurate neutrino energy reconstruction
using only the measured energy and angle of the outgoing lepton. 

In this report, we will cover \ccqe measurements released since
Neutrino 2008, by MiniBooNE, SciBooNE, MINOS and NOMAD.  

\subsection{A few words on theory}

The neutrino-nucleon \ccqe scattering cross-section is most commonly
written according to the Llewellyn-Smith prescription~\cite{Llewellyn
Smith:1971zm}, which parameterises the cross section in terms of
several form factors that are functions of the square of the
four-momentum transferred to the nucleon, $Q^2=-({\bf p}_\nu-{\bf
p}_{\mu})^2$.  Many of the form factors can be taken from electron
scattering experiments.  However, the axial form factor can best be
measured at non-zero \q2 \, in neutrino scattering.  Most experiments
assume a dipole form for the axial form factor $F_A$, such that
$F_A(Q^2)= F_A(Q^2=0)/(1+Q^2/(M_A^{QE})^2)^2$, and use reconstructed
\q2 \, distributions to extract a value for the axial mass parameter
\maqe.

To approximate the nuclear environment, the relativistic Fermi gas
(RFG) model of Smith and Moniz is used by most
experiments~\cite{Smith:1972xh}.  This model assumes that nucleons are
quasi-free, with an average binding energy and Fermi momentum specific
to the particular target nucleus.  Pauli blocking is included in the
model.  Bodek and Ritchie's extension to the relativistic Fermi gas
model \cite{Bodek:1980ar} is employed by some experiments.

These models are predicated on the impulse approximation, which
assumes that the neutrino nucleus interaction can be treated as an
incoherent sum of scattering processes with the individual nucleons.
While such simple models have been demonstrated inadequate for
electron scattering experiments, previous neutrino scattering
measurements were not sufficicient to demonstrate model deficiencies.

More details of the theory of neutrino-nucleus scattering, and
especially progress in new models, are discussed elsewhere in these
proceedings~\cite{AlvarezRuso:nu2010,Benhar:nu2010}.

\section{Neutrino Beam Flux Predictions}

\begin{figure}
\center
{\includegraphics[width=\columnwidth]{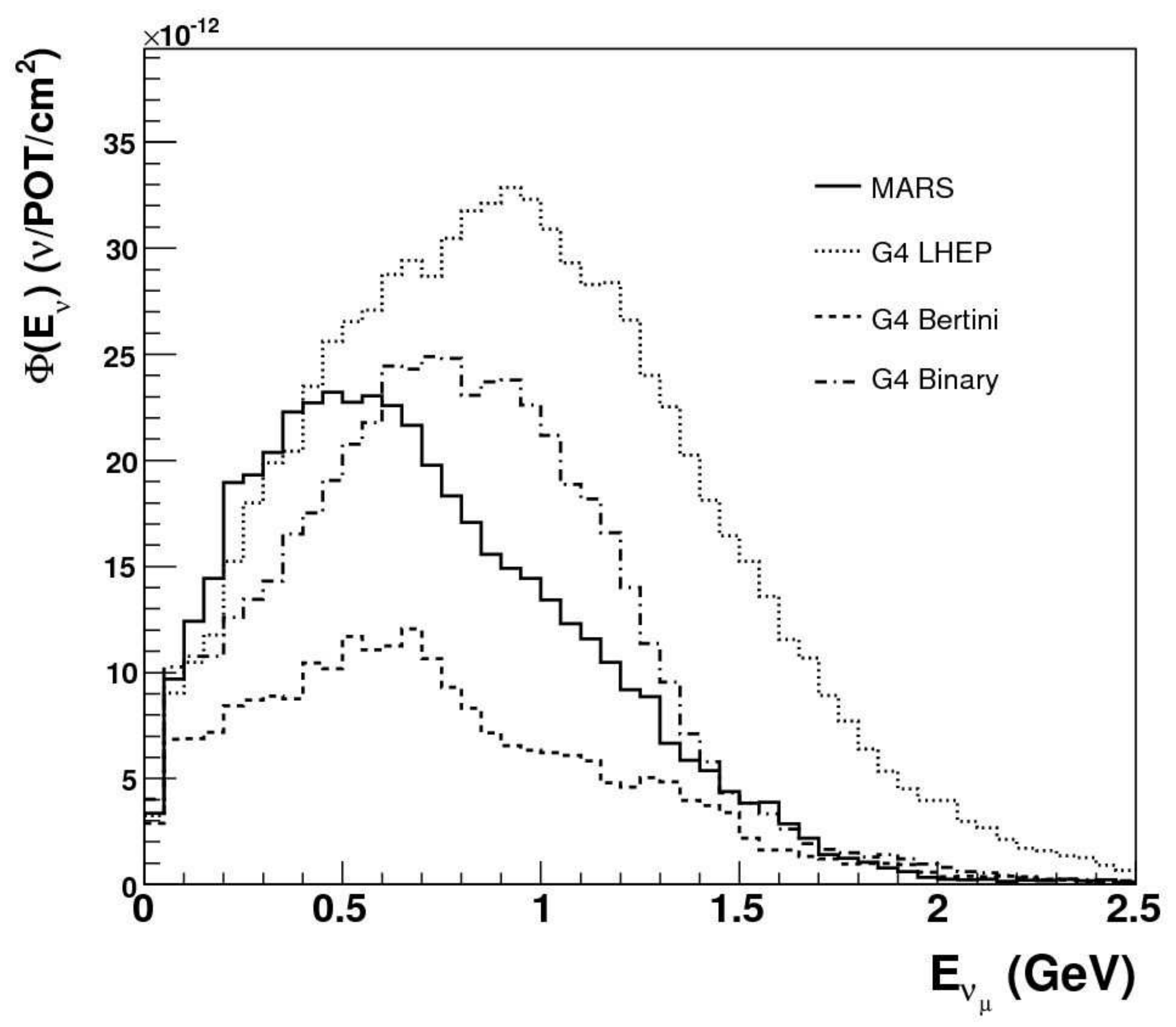}}
\caption{\em {\bf Importance of hadron production for neutrino beams.}
  Four estimates of the neutrino flux at MiniBooNE, using different
  models for the production of parent-pions by p-Be collisions in the
  neutrino target.\cite{Schmitz:2008zz}}
\label{fig:hadron_flux}
\end{figure}

Neutrino cross-section measurements require estimates of the
neutrino fluxes; these estimates have proven to be extremely difficult
since the advent of accelerator neutrino beams.  Most previous
experiments perform some calculations of neutrino fluxes based on
estimates of the secondary pion spectra; these estimates in the past
have had extremely high uncertainties.  Because of this, many past
experiments employed a circular bootstrapping method of estimating the
fluxes.

To illustrate the difficulty of estimating neutrino fluxes,
figure~\ref{fig:hadron_flux} shows four examples of predicted neutrino
flux spectra at the MiniBooNE detector~\cite{Schmitz:2008zz}.  Each flux
prediction was produced using exactly the same Monte Carlo (MC)
simulation of the neutrino target, horn, and secondary beamline, with
the only difference being the primary pion production in each.  The
largest flux estimate is a factor of four higher than the lowest,
illustrating the problem in rather dramatic fashion.

Because of the importance of accurate neutrino flux predictions for
precise cross-section measurements, several experiments have been
performed and planned to make accurate measurements of primary hadron
production cross-sections.  All the measurements discussed herein
use detailed neutrino flux predictions based on precise hadron
production data and/or secondary beam measurements.
Table~\ref{tab:hadprod} summarises the beamline characteristics for
the accelerator neutrino beams used to make the measurements in this
report.

\begin{table}
\begin{center}
\begin{tabular}{c c c c}
\hline \hline

                                & WANF \cite{Astier:2003rj} & NuMI \cite{Kopp:2007zz} & BNB \cite{AguilarArevalo:2008yp} \\ \hline
$E_p (\mathrm{GeV})$            & 450       & 120        & 8     \\
target                          & Be        & C          & Be    \\
$\langle\delta\Phi/\Phi\rangle$ & 7\%       & $\sim$20\% & 9\%   \\
$E_{\nu}$ range                 & 3-100     & 1-20       & 0.2-3 \\
$\langle E_{\nu}\rangle$        & 24.3      & 4          & 0.8   \\
Hadron                          & NA20,SPY  & MIPP       & HARP  \\
prod. exp.                      & (CERN)    & (FNAL)     & (CERN)\\
\hline \hline
\end{tabular}
\end{center}
\caption{\label{tab:hadprod}{\em {\bf Accelerator neutrino beam characteristics}}}
\end{table}

Neutrino flux predictions~\cite{Bishai:nu2010,Kopp:nu2010} and hadron
production experiments~\cite{Blondel:nu2010} are covered in more
detail elsewhere in these proceedings.

\section{Charged-current quasi-eastic measurements}

\subsection{Axial mass measurements}

Due to the paucity of precise neutrino \ccqe \, data, most past
experiments have chosen not to extract the shape of the axial form
factor itself but instead simply find a value of the axial mass that
best fits their data under the assumption of the dipole form.  Here we
report recent fits for \maqe .

MiniBooNE is an 800~t open volume Cherenkov detector.  The MiniBooNE
\ccqe analysis~\cite{AguilarArevalo:2010zc} begins by selecting clean
muon neutrino events, which are identified by observing the muon's
Cherenkov ring followed by the Cherenkov ring produced by the decay
electron.  Requiring the decay electron be located near the end of the
reconstructed muon track yields a high purity \numu \ccqe sample.
Using the full neutrino data set, MiniBooNE finds more than 140,000
events in the \ccqe sample after cuts; this is by far the largest data
set recorded at these energies. The largest fraction of background
events are charged current single pion (CC~$1\pi^+$), \numuccpip,
interactions in which the final state pion is not observed.  This
background is constrained with a sample of \ccpip events selected from
data by tagging events with two decay
electrons~\cite{AguilarArevalo:2009eb,AguilarArevalo:2010bm}.

\begin{table}
\begin{center}
\begin{tabular}{c c}
\hline \hline
Experiment          & $M_A^{QE}$ value \gev2 \\ \hline
World average (d)   & 1.02$\pm$0.03 \cite{Bodek:2007ym} \\
K2K SciFi (O)       & 1.20$\pm$0.12 \cite{Gran:2006jn} \\
K2K SciBar (C)      & 1.14$\pm$0.10 \cite{Espinal:2007zz} \\
MiniBooNE (C)       & 1.35$\pm$0.17 \cite{AguilarArevalo:2010zc} \\
MINOS (Fe)          & 1.19$\pm$0.17 \cite{Dorman:2009zz} \\
NOMAD (C)           & 1.05$\pm$0.06 \cite{Lyubushkin:2008pe} \\
\hline \hline
\end{tabular}
\end{center}
\caption{\label{tab:m_a}\em {\bf Axial mass measurements.} The first row
shows the world average value of $M_A^{QE}$ found by fitting to
previous $\nu_{\mu}-d$ scattering experiments.  The measurements by
K2K, MiniBooNE and MINOS on nuclear targets all use neutrinos in the
few-GeV region, while the NOMAD result ranges from 4-100~GeV neutrino
energy.}
\end{table}

MiniBooNE analysers find that two-dimensional plots of the cosine of
the muon angle versus the muon kinetic energy disagree with their
Monte Carlo (MC) simulation.  Furthermore, they find that the
discrepancy follows lines of constant \q2 , not lines of constant \enu
, which suggests that the source of the disagreement lay with the cross
section model, not the neutrino flux prediction. Based on shape-only
comparisons, the MiniBooNE data show reduced production at low \q2 \,
(below $\sim$0.1~\gev2 ) and increased production above that.  By
fitting the reconstructed \q2 \,distribution MiniBooNE finds the value
of \maqe to be 1.35$\pm$0.17~\gev2 .  The high value of \maqe corrects
the discrepancies in the \q2 \,distribution and improves the
normalization agreement between data and MC.

The MINOS near detector is a 980~t iron calorimeter with a $\sim$~1~T
toroidal magnetic field.  Combined with the intense flux of the NuMI
beam the near detector has recorded an enormous neutrino data set.
For their \ccqe measurement, MINOS analysers select \numu CC events
with low hadronic shower energy.  Similar to MiniBooNE, they find
their data show a deficit compared to their MC simulation at low \q2
(below $\sim$0.1~\gev2 ) but prefer a flatter spectrum above that.
They perform fits of their reconstructed \q2 \, distributions and extract
a value of $M_A^{QE}=1.19\pm0.17$~GeV/c$^2$\cite{Dorman:2009zz} at
mean neutrino energy 3~GeV.  MINOS analysers are currently working on
fits that use non-dipole form factors and developing methods for
constraining the non-QE backgrounds with data.

The NOMAD detector comprised 2.7~t of active drift chamber targets
inside a $0.4$~T dipole magnet.  The excellent resolution provided by
the drift chambers allows the analysers to make stringent cuts on
particle identification parameters and final state configurations.
They select 1-track and 2-track (\mup) \numu \ccqe event samples.  The
NOMAD analysis~\cite{Lyubushkin:2008pe} proceeds by using the measured
yield of deep inelastic scattering (DIS) events, which have a well
known cross-section at high energy, to convert the measured yield of
\ccqe events into a \ccqe cross-section measurement in bins of
neutrino energy.  The measured value of the cross-section in each bin
is used to infer the value of \maqe .  As cross checks, they also fit
the shape of the reconstructed \q2 \, to extract \maqe \, and use the
yield of inverse muon decay (IMD) events to normalize the \ccqe
cross-section.  Both cross checks produce consistent values of \maqe .

Table~\ref{tab:m_a} summarises the recent measurements of \maqe.  We
see that the measurements on nuclei in the $\sim$1~GeV region are
significantly higher than the world average taken from
neutrino-deuterium scattering, and also larger than the high energy
measurement on carbon made by NOMAD.  We stress that the MiniBooNE and
MINOS results (as well as the previously published K2K results) are
based on fits to the shapes of the \q2 \, distributions.

\subsection{Cross-section versus neutrino energy}

\begin{figure*}
\center
{\includegraphics[width=2\columnwidth]{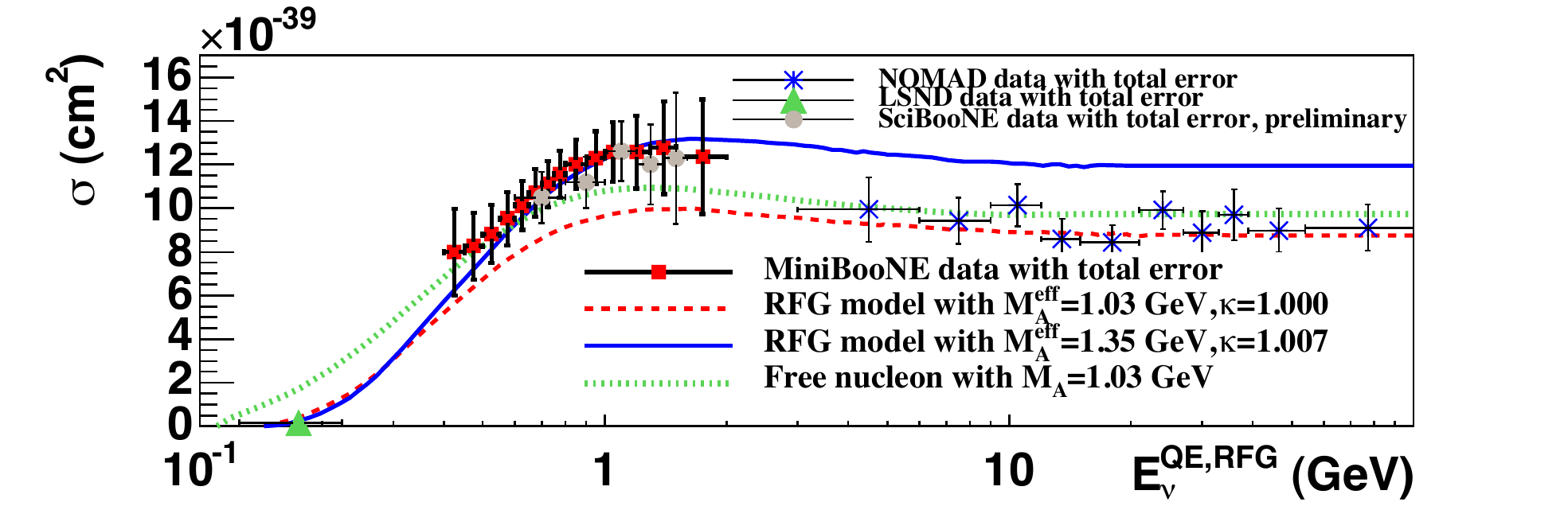}
\caption{\em {\bf \ccqe cross-section versus neutrino
energy.~\cite{AguilarArevalo:2010zc}} The measurements made near 1~GeV
from MiniBooNE and SciBooNE are $\sim$30\% higher in normalization
than what would be expected for a value of \maqe \, consistent with
the world average from Table~\ref{tab:m_a}.  However, the NOMAD
results at higher energies agree well with that expectation.}
\label{fig:enu}}
\end{figure*}

SciBooNE uses its 15~t fine-grained plastic scintillator vertex
detector (SciBar) in combination with its muon range detector (MRD)
for its \ccqe analysis.  Charged-current neutrino candidates are
selected by matching tracks originating in the fiducial volume of
SciBar and penetrating into the MRD; the muons are tagged by their
penetration into the MRD.  The analysers separate events based on the
number of tracks coming out of the neutrino interaction vertex.  One
track events have no tracks other than the muon candidate.  Two track
events are separated into \mup and \mupi samples using particle
identification based on the energy deposited along the second track.
The one-track and \mup samples are predominantly \ccqe events, and the
\mupi sample is predominantly \ccpip events so the analysis constrains
the background fraction with data.  SciBooNE fits reconstructed
$p_{\mu}-\theta_{\mu}$ distributions in the three data samples (single
$\mu$, \mup \, and \mupi ) simultaneously to extract the \ccqe cross
section versus neutrino energy~\cite{AlcarazAunion:2009ku}.

As mentioned above, NOMAD actually directly measures the \ccqe \,
cross-section as a function of neutrino energy, not \maqe .  MiniBooNE
measures the cross-section as a function of neutrino energy separately
from the \q2 \, shape fits used to extract \maqe .

Figure~\ref{fig:enu} compares the measured \ccqe cross-sections versus
neutrino energy from these experiments.  It can be seen that the
MiniBooNE and SciBooNE results, near 1~GeV, are consistent with each
other and are significantly higher than the NOMAD results which span
3-100~GeV.  We note that the SciBooNE and MiniBooNE results are
obtained directly from the measured event yields and
proton-on-target-normalized neutrino flux predictions, not by
extracting via a cross-section ratio with a different (assumed)
cross-section.  These are the world's first such POT-normalized
neutrino \ccqe cross-section measurements.

\subsection{MiniBooNE's differential cross-section}

\begin{figure}
\center
{\includegraphics[width=\columnwidth]{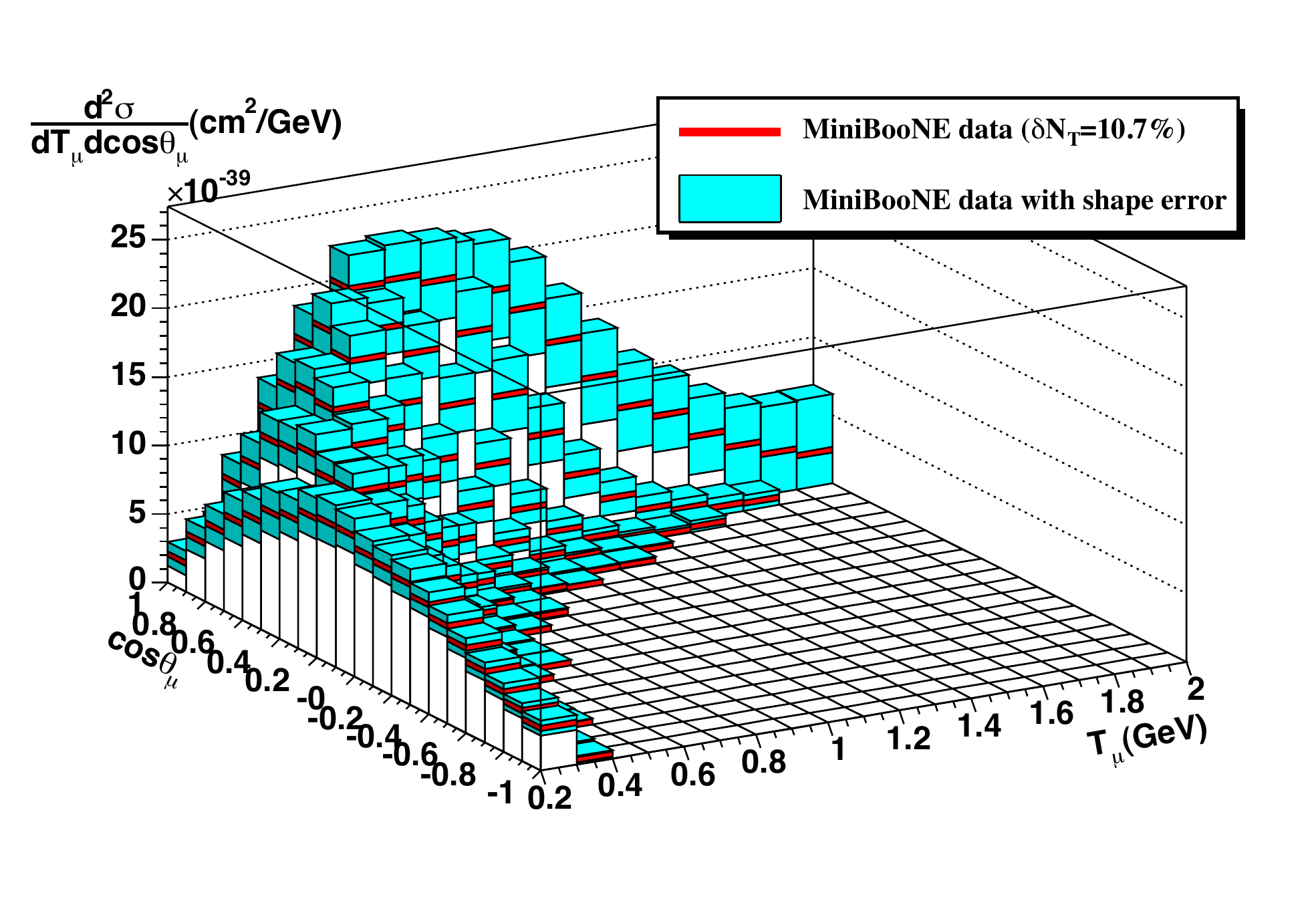}}
\caption{\em {\bf MiniBooNE double differential \ccqe measurements.}
The error bars shown are the shape uncertainties; there is also a
10.7\% normalization error.\cite{AguilarArevalo:2010zc}}
\label{fig:pmu-thetamu}
\end{figure}

Due to the theoretical uncertainties associated with modeling the
nuclear environment, the previous measurements necessarily have some
model-dependence.  To allow for the development of new models,
measurements with no model dependence are needed.
Figure~\ref{fig:pmu-thetamu} shows MiniBooNE's flux-averaged double
differential \ccqe\, cross-section measurement,
$d^2\!\sigma/dT_{\mu}\,d\!\cos\theta_{\mu}$.  This is the most
complete information about the neutrino \ccqe \, cross-section that
can be obtained using the outgoing muon's kinematics.  MiniBooNE has
been on the frontier of model-independent measurements and we hope
that future experiments will follow the example.

\section{Discussion}

\subsection{Increased \ccqe \, cross-section}

Because of the dipole form assumed for $F_A$, changing $M_A^{QE}$
from $1.0$~\gev2 \, to $1.2$~\gev2 \, changes not only the \q2 \,
spectrum but also increases the normalization by approximately 20\%.
The MiniBooNE data actually favor an additional normalization increase
of $\sim8\%$ on top of the larger than expected normalization implied
by the high value of $M_A^{QE}$.  As shown in Figure~\ref{fig:enu},
the preliminary SciBooNE \ccqe analysis is consistent with increased
cross-section at neutrino energy near 1~GeV.  In contrast, the NOMAD
measurement shows good agreement with the normalization expected from
a value of \maqe \, near 1.0 \gev2 .

There is some reason to believe that the impulse approximation may be
inadequate in the 1~GeV region.  Some recent theory papers predict an
increased cross-section in medium sized nuclei~\cite{Martini:2009uj}
near 1~GeV.  Other attempts to fit the MiniBooNE double differential
\ccqe \, data indicate that new ideas are needed~\cite{Butkevich:2010cr,
Benhar:2010nx,Juszczak:2010ve}.  More discussion of this topic can be
found elsewhere in these
proceedings~\cite{AlvarezRuso:nu2010,Benhar:nu2010}.

\subsection{Behavior at low $Q^2$}
\label{sec:lowq2}

In addition to the harder $Q^2$ spectra observed by recent experiments
with \enu \, near 1~GeV (which lead to the high values of $M_A^{QE}$),
$\nu_{\mu}$ \ccqe data have also shown suppressed cross-section at very
low $Q^2$. Several different approaches to dealing with his have been
employed by recent experiments.  For example, the MiniBooNE
collaboration inserted a scale factor into the RFG model that expands
the available phase space for Pauli
blocking~\cite{AguilarArevalo:2007ru} and reduces the predicted
interaction rate at low $Q^2$.  The MINOS collaboration achieves a
similar effect by scaling the Fermi momentum for \ccqe events with
$Q^2 < 0.3$~(GeV/c)$^2$~\cite{Dorman:2009zz}.

In their differential \ccqe analysis, MiniBooNE largely mitigated the
low $Q^2$ data deficit by constraining the \ccpip backgrounds with
their own data.  This indicates the problem lies with the model for
predicting background events, and the observation is supported by
SciBooNE~\cite{Walding:2010zz,AlcarazAunion:2010thesis} and
MINOS~\cite{Mayer:2011nuint11} data.  New and better measurements of \ccpip
production are therefore needed to understand \ccqe scattering; recent
\ccpip measurements are discussed in some detail elsewhere in these
proceedings~\cite{Tzanov:nu2010}.

We note that many in the theory community blame this low $Q^2$
model/data discrepancy on the simplicity of the RFG nuclear model
employed by most experiments.  In this $Q^2$ region, the wavelength of
the boson propagator exceeds the size of the nucleon and so the
assumptions of the impulse approximation break
down~\cite{Ankowski:2010yh}.

\section{Conclusion}

To meet the demands of the current, and future, generation of neutrino
oscillation experiments, serious effort is being put into improved
measurements of neutrino scattering cross-sections.  We have shown
that recent measurements of the \numu \, \ccqe \, process on nuclei
near 1~GeV independently show significant disagreements with the
traditional models of neutrino scattering.  Interestingly, recent
measurements at high energy do not show the same disagreements.  To
solve this puzzle, there is increasing consensus that the neutrino
scattering community needs to provide model-independent measurements
of neutrino cross-sections.  MiniBooNE has led the way in developing
such analyses with SciBooNE following suit now as well.  We look
forward to new data from MINER$\nu$A, Argoneut, T2K and MicroBooNE to
help us solve this latest neutrino conundrum.


\end{document}